\documentclass[conference]{IEEEconf}
\input epsf
\usepackage{graphicx}

\usepackage{pifont}
\usepackage{marvosym}
\usepackage{cite}
\usepackage{amsmath,amssymb,amsfonts}
\usepackage{algorithmic}
\usepackage{textcomp}
\usepackage{xcolor}
\usepackage{colortbl}
\usepackage{multicol}
\usepackage{multirow}
\usepackage{makecell}
\usepackage{mathtools}
\usepackage{listings}
\usepackage{subcaption}
\usepackage{xspace}

\newcommand{\sys}{{\textit{Nimbus}}\xspace}

\def\BibTeX{{\rm B\kern-.05em{\sc i\kern-.025em b}\kern-.08em
    T\kern-.1667em\lower.7ex\hbox{E}\kern-.125emX}}
\begin{document}

\colorlet{mgray}{gray!90!black}

\definecolor{dkgreen}{rgb}{0,0.6,0}
\definecolor{gray}{rgb}{0.5,0.5,0.5}
\definecolor{mauve}{rgb}{0.58,0,0.82}
\definecolor{codegreen}{rgb}{0,0.6,0}
\definecolor{codegray}{rgb}{0.5,0.5,0.5}
\definecolor{codepurple}{rgb}{0.58,0,0.82}
\definecolor{backcolour}{rgb}{0.95,0.95,0.92}

\definecolor{cadmiumgreen}{rgb}{0.0, 0.42, 0.24}

\lstset{
 frame   = tb,
 aboveskip  = 2mm,
 belowskip  = 2mm,
 showstringspaces = false,
 showtabs  = false, 
 columns   = flexible,
    basicstyle=\fontencoding{T1}\footnotesize\fontfamily{lmtt}\selectfont,
 numbers   = left,
 stepnumber  = 1, 
 numbersep  = 1em, 
 numberstyle  = \scriptsize\color{cyan},
 keywordstyle = \color{cadmiumgreen},
 commentstyle = \color{mgray},
 stringstyle  = \color{violet},
 breaklines  = true,
 breakatwhitespace = true,
 tabsize   = 3,
 captionpos  = b, 
 breaklines  = true, 
 breakatwhitespace = true, 
}

\lstdefinelanguage
   [x64]{Assembler}     
   [x86masm]{Assembler} 
   {
        morekeywords={CDQE,CQO,CMPSQ,CMPXCHG16B,JRCXZ,LODSQ,MOVSXD, movabs, movl, movsbl, %
                  POPFQ,PUSHFQ,SCASQ,STOSQ,IRETQ,RDTSCP,SWAPGS,CALLQ,JMPQ,MOVZBL,MOVSLQ,NOPL,NOPW,CMPQ,RETQ, %
                  rax,rdx,rcx,rbx,rsi,rdi,rsp,rbp,rip, %
                  r8,r8d,r8w,r8b,r9,r9d,r9w,r9b, %
                  r10,r10d,r10w,r10b,r11,r11d,r11w,r11b, %
                  r12,r12d,r12w,r12b,r13,r13d,r13w,r13b, %
                  r14,r14d,r14w,r14b,r15,r15d,r15w,r15b, cmpl, endbr64, vext.8, vpmull.64, veor, vpmull2.p64, vpmull.p64, .word, .bbInfo_FUNB, .bbInfo_BB, .bbInfo_BE, .bbInfo_FUNE, .bbInfo_JMPTBL, .quad, xorl, pushq, lui, daddu, daddiu, ld, jr},
    } 

\title{\textbf{\Large \sys: Toward Speed Up Function Signature Recovery via Input Resizing and Multi-Task Learning\\}}

\author{Yi Qian\IEEEauthorrefmark{4}\IEEEauthorrefmark{3}\IEEEauthorrefmark{2}, Ligeng Chen\IEEEauthorrefmark{4}\IEEEauthorrefmark{3}\IEEEauthorrefmark{2}\Letter, Yuyang Wang\IEEEauthorrefmark{4}\IEEEauthorrefmark{3}, and Bing Mao\IEEEauthorrefmark{4}\IEEEauthorrefmark{3}\\
    \IEEEauthorrefmark{4}State Key Laboratory for Novel Software Technology, Nanjing University, Nanjing, China\\
    \IEEEauthorrefmark{3}Department of Computer Science and Technology, Nanjing University, Nanjing, China\\
	\normalsize \{yi\_qian, chenlg, y.y.wang\}@smail.nju.edu.cn maobing@nju.edu.cn\\ \IEEEauthorrefmark{2}Equal Contribution
}



\maketitle
\begin{abstract}
Function signature recovery is important for many binary analysis tasks such as control-flow integrity enforcement, clone detection, and bug finding. Existing works try to substitute learning-based methods with rule-based methods to reduce human effort.They made considerable efforts to enhance the system's performance, which also bring the side effect of higher resource consumption. However, recovering the function signature is more about providing information for subsequent tasks, and both efficiency and performance are significant.

In this paper, we first propose a method called \sys for efficient function signature recovery that furthest reduces the whole-process resource consumption without performance loss. 
Thanks to information bias and task relation (i.e., the relation between parameter count and parameter type recovery), we utilize selective inputs and introduce multi-task learning (MTL) structure for function signature recovery to reduce computational resource consumption, and fully leverage mutual information. Our experimental results show that, with only about the one-eighth processing time of the state-of-the-art method, we even achieve about 1\% more prediction accuracy over all function signature recovery tasks. 
\end{abstract}
\IEEEoverridecommandlockouts
\begin{keywords}
\itshape Function signature; multi-task learning; recurrent neural network
\end{keywords}
\IEEEpeerreviewmaketitle

\section{Introduction}
Function signature recovery plays an important role in binary analysis, widely used in many security analysis works as pre-processing such as bug finding \cite{song2008bitblaze, saxena2009loop}, clone detection \cite{saebjornsen2009detecting, hemel2011finding}, code hardening \cite{wartell2012securing, zhang2013control, zhang2013practical, prakash2015vfguard, van2016tough, muntean2018tau, lin2019control}, etc. 
It is composed of two tasks, parameter count recovery, and parameter type recovery. 

\textbf{Existing Works.}
Function signature recovery is challenging work since most binaries in real applications are stripped, which lose almost all high-level semantics and retain only low-level information via machine code. 

The majority of previous works mainly rely on experienced analysts to recover the missing semantics from binary code \cite{balakrishnan2007divine,caballero2009binary,lin2010automatic,lee2011tie,elwazeer2013scalable,van2016tough,muntean2018tau}.
Recently, some researchers leverage machine learning-based methods to avoid extracting excessive rules via much human effort.
EKLAVYA \cite{chua2017neural} and ReSIL \cite{lin2022resil} utilize gated recurrent unit (abbreviated as GRU) \cite{cho2014properties} (one kind of recurrent neural network) and get surprising results.
Coincidentally, both of them focus on improving recovery performance (i.e., accuracy),
but they leave the problem of \textit{resource consumption} aside, which should be taken into consideration for production.

\textbf{Our Solution.}
In this work, we take all the advantages of machine learning models and try to optimize resource consumption from the whole lifecycle of tool construction. We introduce the 2 key designs to construct our efficient function signature recovery tool \sys\footnote{The name of our system is derived from the book \textit{``Harry Potter"}, and the name is taken from the main character's broomstick \textit{Nimbus 2000}.} as follows, i.e., input reduction and multi-task learning.

\ding{182}\textbf{\textit{Reduce the input size via information bias.}} According to our empirical study of a considerable dataset, we find that the information about function signature is mainly gathered in the front of a function rather than uniformly distributed throughout the function (no matter in binary or assembly level).
Taking the input from the function head achieves more precise and faster performance growth in all function signature recovery tasks than inputting it all into the procedure.

\ding{183}\textbf{\textit{Merging the learning models via mutual information of different tasks.}}
Intuitively, the data distribution of parameter count and parameter type are relevant to each other as well as the function semantics. According to our data analysis, distribution relations widely exist. 
Existing works treat each function signature recovery task independently. Specifically, they recover each function signature (amount or type) with an independent model, called single-task learning (STL) structure. This not only requires independent models to perform prediction tasks separately, consuming a lot of memory and running time but also ignores the mutual information between task associations.

So we introduce multi-task learning (MTL) \cite{caruana1997multitask} structure, which enables deep learning models to train on multiple related tasks on one model and eventually get multiple outputs for different tasks. MTL avoids repetitive work compared with the STL structure, saving resources in both training and testing procedures. In addition, the MTL structure fully utilizes the related information via recovery tasks, improving the performance and the ability to generalization.

\textbf{Evaluation.}
We set up a credible dataset following the previous works, and thoroughly evaluate our system \sys on different aspects. Compared with the state-of-the-art method EKLAVYA, \sys is \textbf{9.92}$\times$ faster on the training procedure and saves about \textbf{87.8\%} time in the prediction procedure both on GPU-equipped and CPU-only hardware environments. Our lightweight and efficient tool design can help security analysts perform pre-analysis or provide analysis results for downstream tasks.

The contributions of this paper are as follows: 
\begin{itemize}
    \item According to the empirical study, we verify the information bias phenomenon in each binary level function, and we prune the functions to keep highly informative instructions as input.
    \item According to the intuitive relations of sub-tasks in function signature recovery, we first introduce the multi-task learning structure to enhance function signature recovery and customize an MTL-GRU architecture. 
    \item We evaluate the prototype of \sys on the dataset, and our work achieves a significant reduction in resource consumption, while even getting about a 1\% accuracy increase.
\end{itemize}

The rest of the paper is organized as follows, Section \ref{motivation} introduces the motivation of our work.
Section \ref{problem_def} defines the problem. 
Section \ref{design} presents the workflow and our system design.
Section \ref{evaluation} evaluates our model.
Section \ref{discussion} presents the discussion. 
Section \ref{related_works} discusses some related works and Section \ref{conclusion} concludes the paper.

\section{motivation}\label{motivation}
Most learning-based function signature recovery works are hard to use as a tool because they require hours or even days of training on high-power GPUs and cost a lot when predicting even though they achieve good performance. 
For example, we rebuild EKLAVYA and find that the average time for predicting a single function using GPU and CPU is \textbf{17.88ms} and \textbf{70.40ms}, respectively. 
As a pre-work for many binary analysis tasks, such resource consumption may affect their efficiency, especially without high-power GPUs. In this paper, we try to reduce the resource consumption of both training and prediction. Nowadays, most works focus on improving model performance, our work balance the performance and overhead. We hope this contributes to the community.



\begin{figure}[!h]
\centerline{\includegraphics[width=0.45\textwidth]{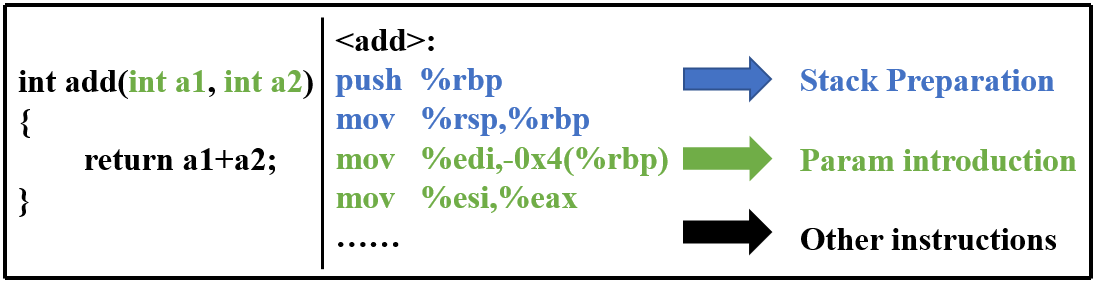}}
\caption{Source code and its assembly code.}
\label{mot_asm}
\end{figure}

\subsection{Model input}

Most existing works use unfixed-length assembly codes or byte codes as model inputs. 
Obviously, the longer the input length, the more information the model obtains, and the easier it is to improve the model performance. However, longer inputs also require more computational resources. By reducing the input length, we can save computational resources from the source. 
By following the data selection mechanism of EKLAVYA, we compile some widely used open-source projects into binaries as our dataset. We analyze the dataset and there are two key findings in the selection of model inputs. The detail of the dataset is given in Section \ref{evaluation}.

\subsubsection{\textbf{Information bias - Does the information in the content of the code follows a uniform distribution}}
During analysis, we find that although the use of parameters is scattered through the whole function, they are often introduced at the function head. Figure~\ref{mot_asm} gives an example function and its assembly code, and the import of parameters clusters at the function head. 

We anticipate that this phenomenon widely exists in binary codes, i.e., that information about the function signature clusters at the function head and we call this \textit{information bias}. 
We do several experiments to verify the existence of information bias in Section \ref{evaluation}.

\subsubsection{\textbf{Input length - What input length is the most appropriate}}

\begin{figure}[!h]
\centerline{\includegraphics[width=0.3\textwidth]{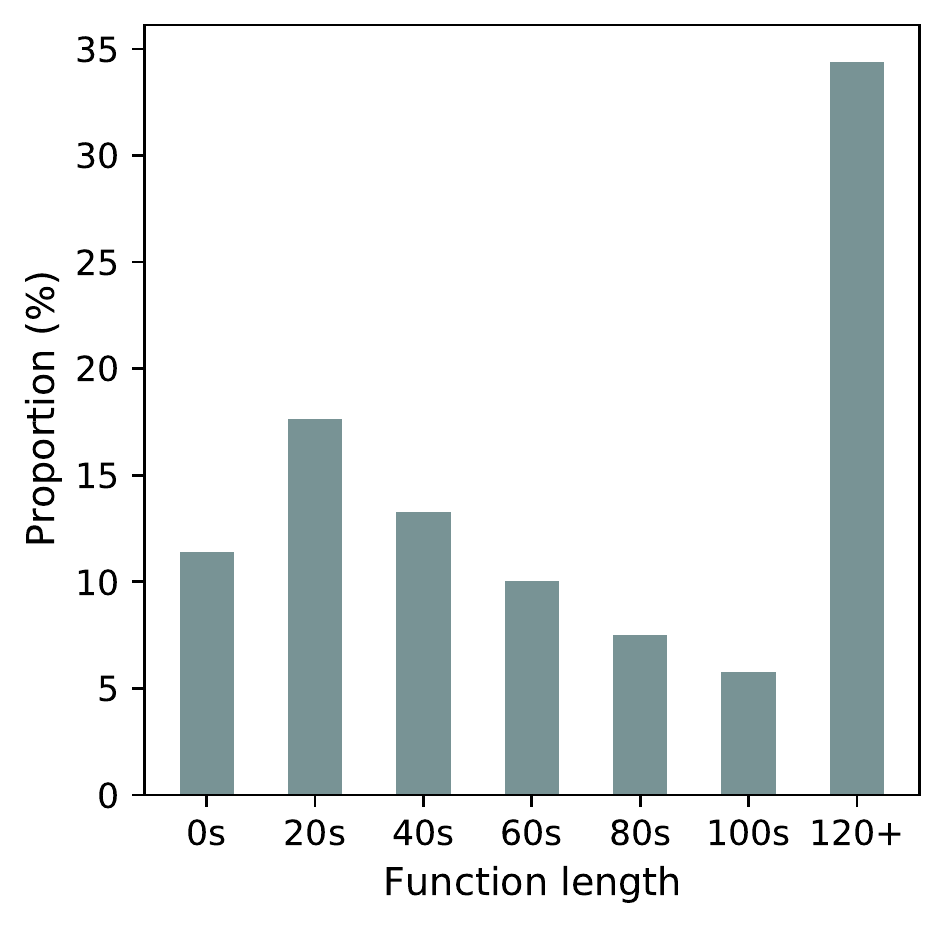}}
\caption{Function lengths and their proportions. '0s' represents the function length is between [0, 20), and so on.}
\label{eval_funclength}
\end{figure}

According to the analysis results of the function length shown in Figure~\ref{eval_funclength}, we find that about 30\% of the functions are less than 40 instructions and about 70\% of the functions are less than 120 instructions, i.e., most functions are not that long.  Due to information bias, when the input reaches a certain length, the following instruction may contain little valuable information, so the overlong inputs may not be required.

Another interesting finding is a large gap between the mean and the median of the function lengths. Precisely, the mean is 165 and the median is 74.
Combined with the statistical distribution, we believe that the median is more representative of the general function length. To observe the effect of different input lengths on the model, we train models with different input lengths around the median in Section \ref{evaluation}.

\subsection{Model structure}
Previous works treat function signature recovery tasks as independent and use STL models to recover them. Specifically, one model can only recover parameter count or one parameter type at one specific location, so multiple models are required to accomplish function signature recovery. 
However, in different models, the same layers undertake often similar work as these tasks are related, which undoubtedly leads to a waste of computational resources. As a result, in the case of limited computational power, it takes multiple times to train and use models. 
If we can avoid repeated computation in those STL models, we can reduce computational resource consumption structurally.

\subsubsection{\textbf{Task relation}}
Function signature recovery consists of two parts: parameter count recovery and parameter type recovery. Intuitively, there are two possible relations between tasks. One is the relation between parameter count and parameter types, and the other is between two parameter types in different positions. 
We try to explore if these relations really exist. And if so, whether we could improve performance with them.

\begin{figure}[!h]
\centerline{\includegraphics[width=0.48\textwidth]{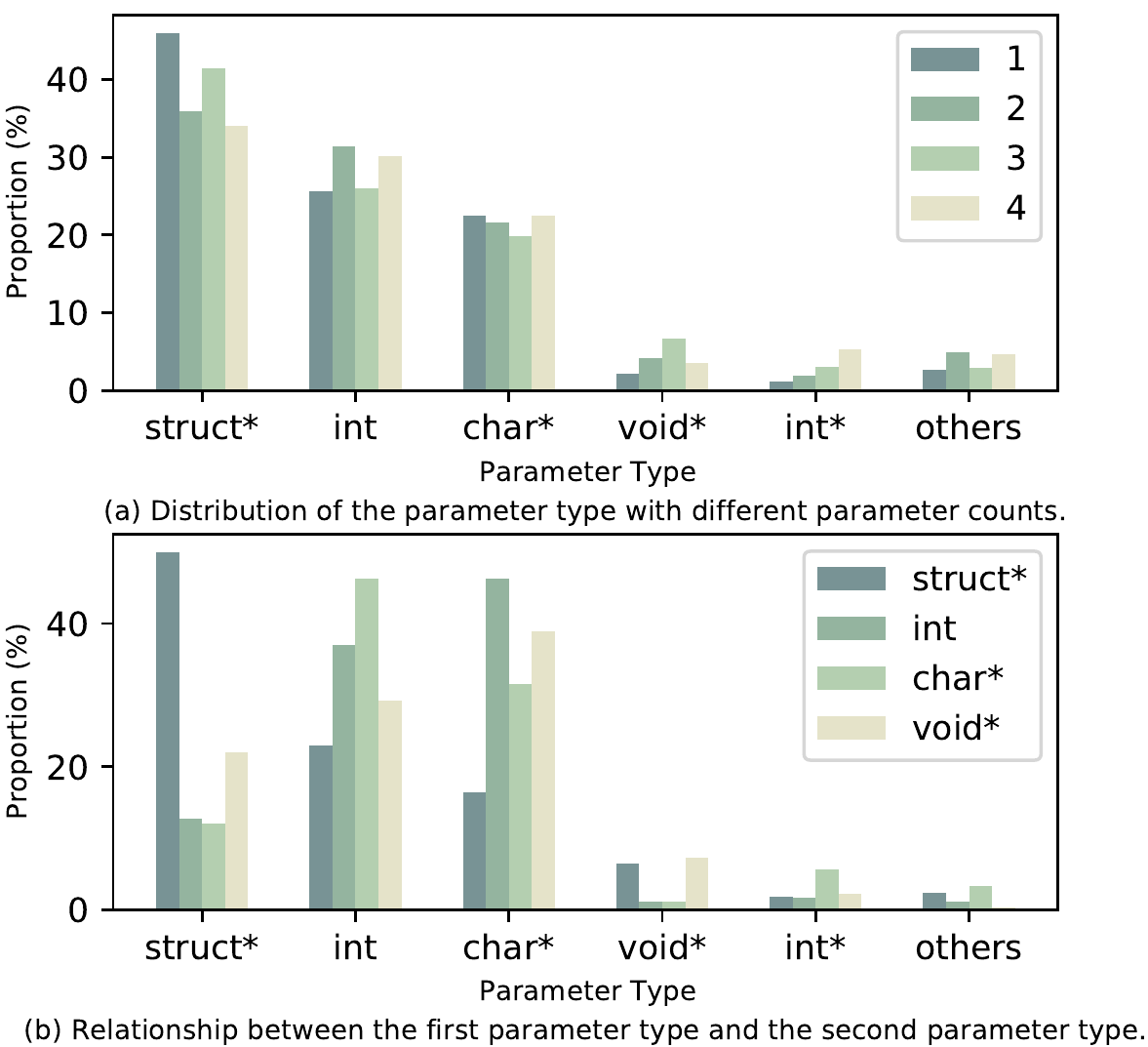}}
\caption{Relations of recover targets. To make the relations clearer we have omitted some types that have a smaller proportion.}
\label{discuss_params}
\end{figure}

To verify the intuition, we analyze the dataset, and the results are shown in Figure~\ref{discuss_params}. 
According to the results, the parameter count and the parameter type are correlated, and so do the parameter types in different positions.
For example, when the first parameter is \textit{struct*}, the second one is most probably \textit{struct*} as well. 
In fact, we find that the above relations widely exist in the dataset.

\subsubsection{\textbf{Multi-task learning structure}}

The MTL structure \cite{caruana1997multitask} is proposed to make the model learn the information between 
related tasks. It consists of shared layers and task-specific layers. Different tasks share their intermediate representation at the shared layer and get task-specific results at the task-specific layer.

\sys benefits a lot from the MTL structure. First, it helps \sys to select features. For a given task, its related tasks give \sys the evidence of which features are useful and help \sys to focus more on them. Second, the MTL helps \sys learns more generalized representations. A model will be overfitting if it learns both data and noise during training. MTL forces \sys to adapt to the noise of different tasks, which reduces overfitting and makes \sys more generalized. 
Last but not the least, MTL actually merges multiple task-specific layers into one shared layer and it obviously avoids repeated computation thus reducing resource consumption.

\section{problem definition}\label{problem_def}
\begin{figure}[!h]
\centerline{\includegraphics[width=0.49\textwidth]{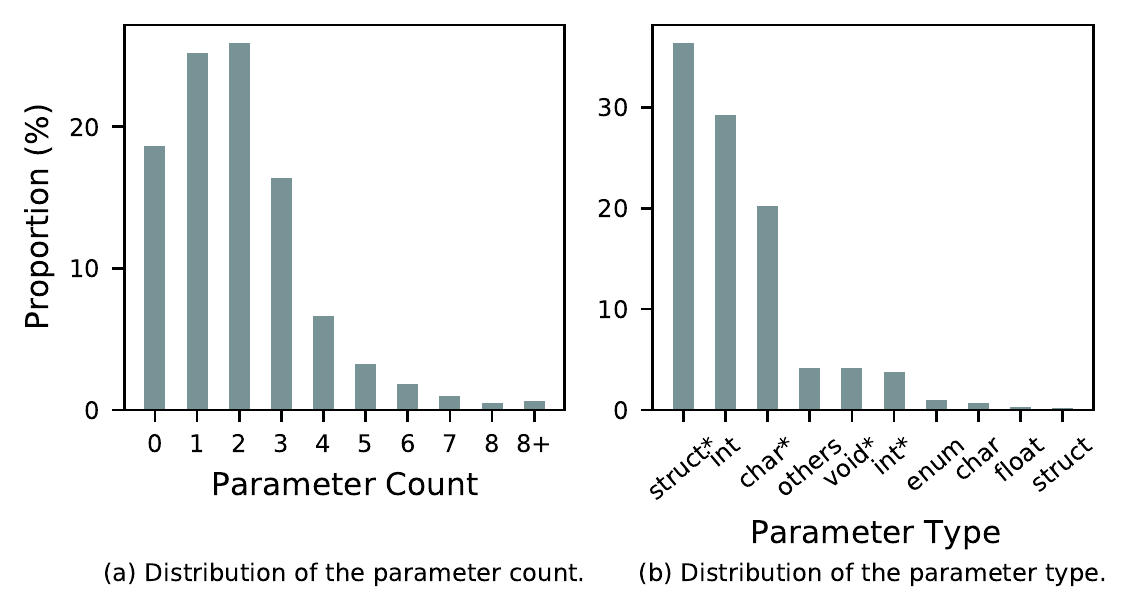}}
\caption{Proportion of parameter count and parameter type. '8+' denotes more than 8.}
\label{def_pc_pt}
\end{figure}

The distribution of parameter counts and parameter types are shown in Figure~\ref{def_pc_pt} (a) and Figure~\ref{def_pc_pt} (b), respectively.
Our function signature recovery tasks are defined as follows. 



\begin{itemize}
\item \textbf{Parameter Count}: The number of parameter passed to function, \textbf{abbreviated as $PC$}.
    \begin{equation}
        PC\in \{0,1,2,3,4,5,6,7,8,others\}
    \end{equation}
    
\item \textbf{Parameter Type}: Parameter type for each parameter passed to function, \textbf{abbreviated as $PT$}. $PT_i$ represents the parameter at the $i_{th}$ position $(i=1,2,...)$.
    \begin{equation}
        \begin{split}
            PT \in \{struct*,int,char*,void*,int*,enum,\\
                            char,void,float,struct,others,NULL\}, 
        \end{split}
    \end{equation}
    
\end{itemize}
where \textit{NULL} denotes that the position has no parameters.

We make such definitions because over 99\% of the function parameters are less than 9, and parameter types defined in $PT$ (except \textit{others} and \textit{NULL}) account for more than 95\% of all parameter types. 
Note that our $PT$ is different from EKLAVYA, we remove the \textit{union} because we think the \textit{union} has more high-level semantics. In assembly code, \textit{union} will be translated into a certain parameter type. In addition, we add other parameter types such as \textit{char*} to make our model suitable for more situations.

The model input is the assembly code from the function that can be easily obtained from the disassemblers, and the output is the function signature $PC$ and $PT$ defined above.
\section{Design}\label{design}
\begin{figure*}[!h]
\centerline{\includegraphics[width=0.75\textwidth]{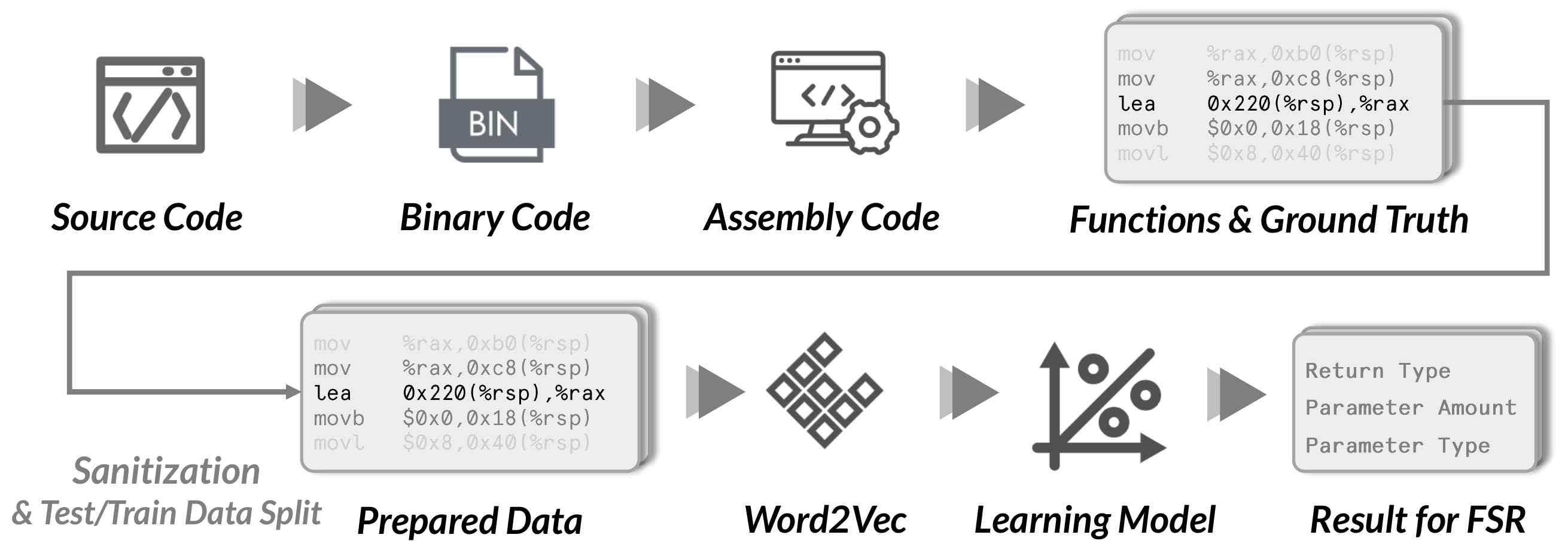}}
\caption{An overview of the steps from building a dataset to recovering function signature.}
\label{design_workflow}
\end{figure*}

The workflow is shown in Figure~\ref{design_workflow}. There are two significant parts to our workflow. One is to vectorize the input (word embedding), and the other is to train the classification model (signature recovery). We make vectorization a separate module because we believe that the mature word embedding techniques retain more semantic information. 
We take assembly code as input, and we employ the MTL structure which allows the model outputs $PC$ and multiple $PT$ at the same time.

\subsection{Word embedding}
To make the input learnable for the model, the first step is to vectorize the input, which is called word embedding. By representing semantically similar text with similar high-dimensional vectors, word embedding essentially reprocesses the input and improves the representation ability.

There are many word embedding techniques, including \textit{one-hot representation}, \textit{word2vec} \cite{mikolov2013distributed}, and \textit{fasttext} \cite{bojanowski2017enriching}. Our experiment employs \textit{word2vec}, which is a mature word embedding tool that vectorizes words quickly and effectively given corpus. 

Notice that we are word embedding the instruction words (mnemonics and operands) instead of the instructions. For example, we split \textit{mov a,b} as \textit{mov}, \textit{a} and \textit{b} and embedding them. Embedding instruction words has two advantages. First, it disperses the semantic information from a single vector to multiple vectors, and the dimension of a single vector can be reduced. Thus the computational consumption can be reduced. The second is that the relations within the instructions can be captured, allowing the model to learn more content details. 
In particular, we always split one instruction into four instruction words and truncate for larger ones because we find that most instructions can be converted into less than four instruction words, which also facilitates our subsequent processing. At last, we use the continuous bag of words (CBOW) negative sampling method in \textit{word2vec} to train instruction words into 128-dimensional vectors.

\subsection{Signature recovery}



\begin{figure}[!h]
\centerline{\includegraphics[width=0.49\textwidth]{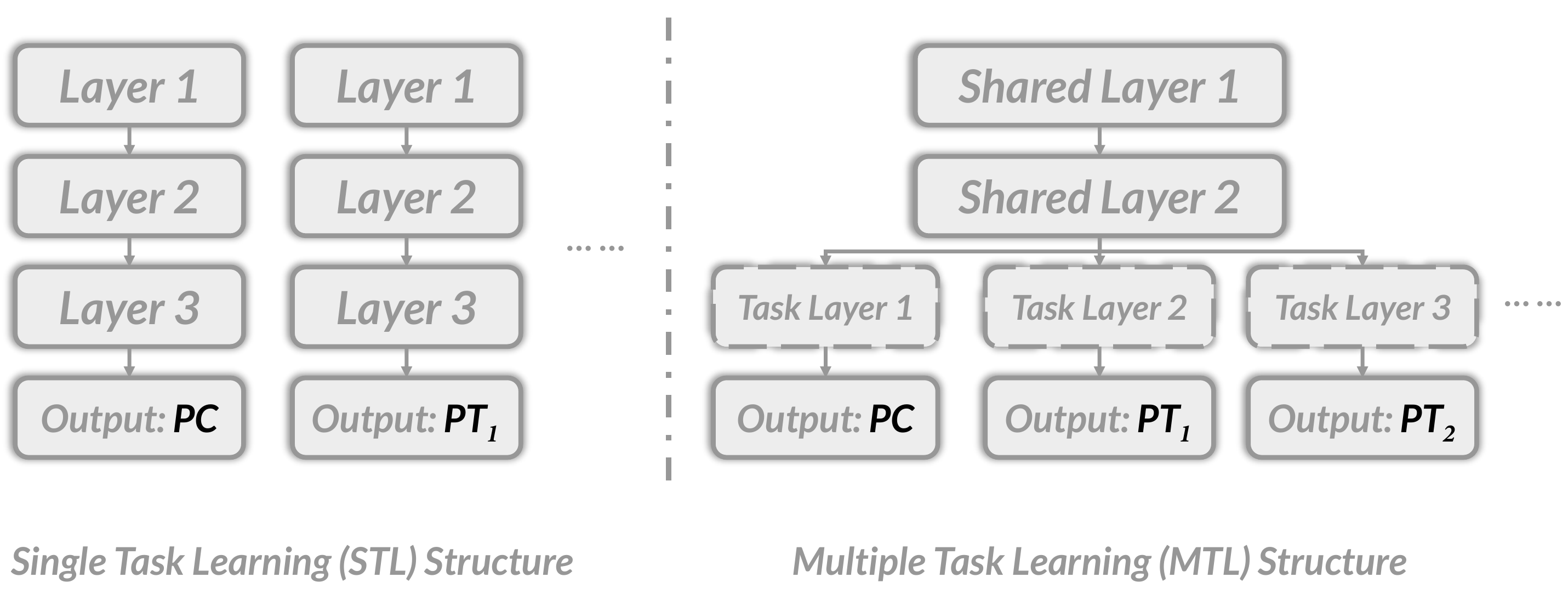}}
\caption{Model structure for recovering $PC$, $PT_1$, $PT_2$, $PT_3$, the model has four outputs.}
\label{design_mtl}
\end{figure}

As we discussed in Section \ref{problem_def}, most functions contains three or less parameters, so we employ a MTL model with one input and four outputs (denoted as $PC$, $PT_1$, $PT_2$ and $PT_3$). 
Our model consists of two shared layers and 4 task-specific layers, the model structure is shown in Figure~\ref{design_mtl}. 
If we want to recover extra parameter types, we only need to add a new task-specified branch rather than a new model.

We choose to use a recurrent neural network (RNN) as the \sys network architecture. 
RNN introduces ``memory" to the model. The network will memorize the previous information and apply it to the calculation of the current output. The ``recurrent" comes from the fact that each node performs the same task. RNN is very effective for sequence data, as it mines time-series information and semantic information in the data. To alleviate the problems of gradient disappearance and gradient explosion, long-short term memory (LSTM) \cite{hochreiter1997long}, a variant of RNN, uses forget gate and input gate to update the previously saved information. And GRU further simplifies the gate structure, merges the forget gate and the input gate into an update gate.
We use GRU because it has fewer parameters and trains faster.

\section{Evaluation} 
\label{evaluation}
In this section, we try to answer the following questions, by evaluating with control variables:
\begin{itemize}
    \item \textbf{RQ1:} How do the \textit{different input lengths} and \textit{resizing strategies} (from head/tail) affect the method performance?
    \item \textbf{RQ2:} How does the \textit{network structure} (MTL/STL) affect the method performance?
    \item \textbf{RQ3:} How much does the optimization of the whole process save \textit{resource consumption}?
\end{itemize}

Our experiments are performed on a server containing one 12-core Ryzen 3900X CPU with 48GB of RAM, and one GeForce RTX 3080 GPU with 10GB of memory. The neural network and data processing routines are written in Python, using Keras \cite{chollet2015keras}.

\subsection{Dataset}
We set up a dataset for method evaluation. The binaries are compiled from source files often used in the community, listed in Table \ref{tab_sourcecode}. 

\subsubsection{Function Extraction}
We compile the source files under different configurations, and the final dataset consists of binary files compiled with different compilers and versions (\textit{clang 3.0, clang 4.0, clang 5.0, clang 6.0, clang 7, gcc 5, gcc 6, gcc 7, gcc 8, gcc 9}), and different optimization levels (\textit{-O0, -O1, -O2, -O3, -Os}) for x64. We use \textit{objdump} \cite{GNU} to disassemble the binary code and get their assembly code in AT\&T format. Finally, we get a dataset consisting of 2,819,495 functions. 

\begin{table}[!h]
\setlength\tabcolsep{15pt}
\caption{Project and version of dataset.}
\begin{center}
\resizebox{0.49\textwidth}{!}{
\begin{tabular}{c|c||c|c}
\Xhline{1.2pt}
\rowcolor{gray}
\textcolor{white}{\textbf{Project}} & \textcolor{white}{\textbf{Version}} & \textcolor{white}{\textbf{Project}} & \textcolor{white}{\textbf{Version}} \\
\Xhline{1.2pt}
Coreutils & 8.31 & Gtypist & 2.9.5 \\
Inetutils & 1.9.4 & Binutils & 2.32 \\
Grep & 3.3 & Gawk & 5.0.0 \\
Nginix & 1.15.12 & Sed & 4.7 \\
Libpng & 1.6.37 & Bash & 5.0 \\
Cflow & - & Less & 530 \\
BC & 1.07 & Bison & 3.4 \\
Nano & 4.4 & Indent & 2.2.12 \\
Wget & 1.20.3 & Gzip & 1.9 \\
\Xhline{1.2pt}
\end{tabular}
}
\end{center}
\label{tab_sourcecode}
\end{table}

\subsubsection{Sanitization and duplication}
To avoid the interference of irrelevant information, we sanitize the specific address and function name. Listing \ref{eval_sanitization} shows a brief example.
As the instructions for direct jump, we replace the concrete address with '\textit{IMM}' (e.g., row 1, 3). As the function reference, we replace the concrete function name with '\textit{FUNC}' (e.g., row 2). 
To avoid repeated functions, we filter the functions in the dataset based on MD5 after sanitization, and only one repeated function is retained to ensure data balance. Only 272,900 distinct functions are retained after sanitization and duplication.

Through compilation, we find that binaries compiled from different projects and optimization levels may the same. To avoid information leaks, sanitization only randomly keeps one of them, which makes the evaluation of different optimization levels biased, so we do not distinguish optimization levels in the following discussion.

\vspace{6pt}

\begin{minipage}[b]{0.95\linewidth}
\lstset{numbersep = 0.5em}
\begin{minipage}[t]{0.45\linewidth}
\begin{lstlisting}[language={[x64]Assembler}]
mov 0x2063a3(%rip), %rsi
je  401f31<add+0x34>
movabs $0xaaaaaaa9, %rax
cmp %rax,%rsi
\end{lstlisting}
\end{minipage}
\hfill
\hfill
\begin{minipage}[t]{0.45\linewidth}
\begin{lstlisting}[language={[x64]Assembler}]
mov IMM(%rip),%rsi
je IMM<FUNC>
movabs IMM,%rax
cmp %rax,%rsi
\end{lstlisting}
\end{minipage}
\captionof{lstlisting}{Assembly code before and after sanitization
\label{eval_sanitization}}
\end{minipage}

We obtain the ground truth for the function signature by analyzing the \textit{DWARF debugging information} \cite{DWARF}. 
We divide the dataset into a training set and a testing set at 8:2.

\subsection{Metrics}
\subsubsection{Performance evaluation}
Precision (\textit{P}) and recall (\textit{R}) are commonly used to evaluate the model performance, which are calculated as
\begin{equation}
    P = \frac{TP}{TP+FP}; R = \frac{TP}{TP+FN},
\end{equation}
where \textit{TP}, \textit{FP}, and \textit{FN} denote true positive, false positive, and false negative, respectively. 



To measure the performance, we use weighted accuracy $Acc$ which can be calculated as
\begin{equation}
    Acc = \sum\limits_{i=1}\limits^{n}\sigma_{i}\times R_{i}, 
\end{equation}
where \textit{n} denotes the number of classes and $\sigma_{i}$ denotes the proportion of the label $i$ in the test set. 

Weighted accuracy represents the correctly predicted rate in the test set which is a widely used metric for multi-classification tasks such as emotion recognition \cite{han2014speech,jin2015speech,akhtar2019multi}, malware classification \cite{raff2018investigation} and text classification \cite{mohtarami2018automatic}. Since weighted accuracy reveals the model's global performance and emphasizes the effect of every label simultaneously, EKLAVYA uses it and so do we.

\subsubsection{Resource consumption evaluation}

Since the computational resource is valuable, we also take optimizing the resource consumption into account besides the model performance. 
We define \textit{Efficiency}:

\begin{equation}
    \begin{split}
        Efficiency = \frac{\sum\limits_{k=1}\limits^{N_t}Acc_k}{\sum\limits_{i=1}\limits^{N_G} T_i \times U_i},
    \end{split}
\end{equation}
where $N_t$ denotes task number, $N_G$ denotes GPU number, $T_i$ denotes time consumption and $U_i$ denotes GPU usage percentage, respectively. The larger of efficiency, the smaller the resource consumed to reach the same accuracy relatively.

\subsection{Experiment on performance}

\subsubsection{\textbf{Compared with EKLAVYA}}
We compare \sys with EKLAVYA in our dataset. We use cross-entropy loss for our classification tasks, and the optimizer is  \textit{Adam} \cite{kingma2015adam} with learning rate of 1e-4, $\beta_1 = 0.9$, $\beta_2 = 0.999$ like many previous works. We use a dropout \cite{srivastava2014dropout} probability of 0.2 on all layers to alleviate overfitting. 
We use 128-dimensional vectors embedded from 40 instructions as input and train for 100 epochs. To make the evaluation fair enough, we reproduce EKLAVYA referring to their paper, with the same hyper-parameters, the results are shown in Table \ref{tab_performance_ek}. \sys performs slightly better than EKLAVYA, we achieve about 1\% more prediction accuracy over all function signature recovery tasks. Column 2 (i.e., \textit{PC}) represents the performance of recovering the amount of function parameters. Column 3 to 5 (i.e., $PT_1, PT_2, PT_3$) represent the performance of recovering variable type of different position parameters, respectively.

\begin{table}[!h]
\setlength\tabcolsep{3.5pt}
\caption{Model accuracy comparison.}
\scriptsize

\begin{center}
\resizebox{0.49\textwidth}{!}{
\begin{tabular}{c|c|c|c|c}
\Xhline{1.2pt}
\rowcolor{gray}
\textcolor{white}{\textbf{Method}} & \textcolor{white}{\textbf{$PC$ (\%)}} & \textcolor{white}{\textbf{$PT_1$ (\%)}} & \textcolor{white}{\textbf{$PT_2$ (\%)}} & \textcolor{white}{\textbf{$PT_3$ (\%)}} \\ \Xhline{1.2pt}
EKLAVYA        & 96.42          & 94.88           & 95.40           & 97.88           \\ 
\textbf{\sys}          &\textbf{ 97.25 (+0.83)}          & \textbf{95.88 (+1.00)} & \textbf{96.82 (+1.42)    }       & \textbf{98.40 (+0.52)}         \\ \Xhline{1.2pt}
\end{tabular}
}
\end{center}
\label{tab_performance_ek}
\end{table}

We further do experiments on inputs and model structures of our model. The settings remain the same if not mentioned. 
We find that MTL benefits our model in performance with appropriate input according to the later experiment.

\subsubsection{\textbf{Information bias shortens the input length}}
We train an STL-GRU model with different instruction lengths and positions, and the results are shown in Table \ref{tab_eval_bias}. The size represents how many instructions are used as input. The location represents where the instructions come from. 

\begin{table}[!h]
\caption{Ablation study of different locations and different instructions' input lengths. (Accuracy)}
\begin{center}
\resizebox{0.49\textwidth}{!}{
\begin{tabular}{c|c|c|c|c|c}
\Xhline{1.2pt}
\rowcolor{gray}
\textcolor{white}{\textbf{Size}} & \textcolor{white}{\textbf{Location}} & \textcolor{white}{\textbf{$PC$ (\%)}} & \textcolor{white}{\textbf{$PT_1$ (\%)}} & \textcolor{white}{\textbf{$PT_2$ (\%)}} & \textcolor{white}{\textbf{$PT_3$ (\%)}} \\
\Xhline{1.2pt}
\multirow{2}{*}{5} & Head & 57.97 & 65.22 & 62.02 & 75.12  \\
  & \textcolor{gray}{Tail} & \textcolor{gray}{47.43} & \textcolor{gray}{57.25} & \textcolor{gray}{55.13} & \textcolor{gray}{74.69}  \\
\hline
\multirow{2}{*}{10} & Head & 89.78 & 85.79 & 86.68 & 91.87  \\
  & \textcolor{gray}{Tail} & \textcolor{gray}{66.17} & \textcolor{gray}{75.02} & \textcolor{gray}{72.41} & \textcolor{gray}{82.74}  \\
\hline
\multirow{2}{*}{20} & Head & 96.23 & 94.23 & 95.15 & 97.25  \\
  & \textcolor{gray}{Tail} & \textcolor{gray}{81.68} & \textcolor{gray}{87.04} & \textcolor{gray}{85.57} & \textcolor{gray}{90.28}  \\
\hline
\multirow{2}{*}{40} & Head & \textbf{96.74} & \textbf{95.46} & \textbf{96.23} & \textbf{97.76}  \\
  & \textcolor{gray}{Tail} & \textcolor{gray}{87.99} & \textcolor{gray}{90.99} & \textcolor{gray}{90.45} & \textcolor{gray}{93.52}  \\
\hline
\multirow{2}{*}{80} & Head & 96.72 & 95.84 & 96.35 & 97.71  \\
  & \textcolor{gray}{Tail} & \textcolor{gray}{92.06} & \textcolor{gray}{93.38} & \textcolor{gray}{93.62} & \textcolor{gray}{95.67}  \\
\hline
\multirow{2}{*}{120} & Head & 97.01 & 95.94 & 96.60 & 97.95 \\
  & \textcolor{gray}{Tail} & \textcolor{gray}{93.86} & \textcolor{gray}{94.63} & \textcolor{gray}{95.00} & \textcolor{gray}{96.50}  \\
\Xhline{1.2pt}

\end{tabular}
}
\end{center}
\label{tab_eval_bias}
\end{table}

When the size is 40 and the input comes from the head and the tail, the accuracy of $PC$ is 96.74\% and 87.99\%, respectively. Experiments also show that 20 instructions from the head achieve an approximately equal performance of 120 from the tail. Actually, instructions from the head always make the model better than that from the tail in any size and any task, which verifies our intuition of information bias. 
In addition, the model performance becomes better as the input length increases. However, there is little accuracy growth since size reaches 40 when the input comes from function head, where the $Acc$ of $PC$, $PT_1$, $PT_2$ and $PT_3$ are 96.74\%, 95.46\%, 96.23\%, and 97.76\% respectively. 
This phenomenon is caused by information bias as well, i.e., little valuable information is contained after the first 40 instructions.

In addition, with the size increases, the accuracy difference between head and tail decreases. A reasonable explanation is that the distribution of function length cause it. As we mentioned in Section \ref{motivation}, about 70\% of functions are less than 120 instructions. In another word, when the size is 120, the instructions in the head are the same as those in the tail for about 70\% of the functions. When the size is large enough, the information obtained from the tail comes from the head, which brings us back to information bias.

To sum up, we think it is best to set the model input as 40 instructions from the head to achieve satisfying results.

\subsubsection{\textbf{MTL structure makes models obtain positive information gain}}
We train STL-GRU and MTL-GRU models with different sizes shown in Table \ref{tab_eval_mtl}. All the input comes from the function head. 
When the size is large enough, MTL performs better than STL in all tasks, e.g., the accuracy of $PC$ is 97.25\% and 96.74\% of MTL and STL with size 40.
This demonstrates the relations between signature recovery tasks, and MTL models obtain information gain and improve generalization to perform better. 
On the contrary, MTL does not perform as well as STL when the size is insufficient, e.g., 20. We conjecture two reasons for this situation. One is information lack, and the other is noise propagation. In case of information lack, STL, focusing on a single task with limited information, is better. Besides, MTL propagates and amplifies the noise between different tasks.
Combined with former experiments, \sys adopts the MTL structure.

\begin{table}[!h]
\caption{Ablation study of model structure. (Accuracy)}
\begin{center}
\resizebox{0.49\textwidth}{!}{
\begin{tabular}{c|c|c|c|c|c}
\Xhline{1.2pt}
\rowcolor{gray}
\textcolor{white}{\textbf{Size}} & \textcolor{white}{\textbf{Structure}} & \textcolor{white}{\textbf{$PC$ (\%)}} & \textcolor{white}{\textbf{$PT_1$(\%)}} & \textcolor{white}{\textbf{$PT_2$ (\%)}} & \textcolor{white}{\textbf{$PT_3$ (\%)}} \\
\Xhline{1.2pt}
\multirow{2}{*}{5} & MTL & 57.59 $\downarrow$ & 65.10 $\downarrow$ & 62.08 $\downarrow$ & 75.00 $\downarrow$ \\
  & \textcolor{gray}{STL} & \textcolor{gray}{57.97} & \textcolor{gray}{65.22} & \textcolor{gray}{62.02} & \textcolor{gray}{75.12} \\
\hline

\multirow{2}{*}{10} & MTL & 89.53 $\downarrow$ & 85.01 $\downarrow$ & 86.10 $\downarrow$ & 91.34 $\downarrow$ \\

 & \textcolor{gray}{STL} & \textcolor{gray}{89.78} & \textcolor{gray}{85.79} & \textcolor{gray}{86.68} & \textcolor{gray}{91.87}   \\
\hline

\multirow{2}{*}{20} & MTL & 95.98 $\downarrow$ & 94.11 $\downarrow$ & 95.44 $\uparrow$ & 97.10 $\downarrow$  \\
  & \textcolor{gray}{STL} & \textcolor{gray}{96.23} & \textcolor{gray}{94.23} & \textcolor{gray}{95.15} & \textcolor{gray}{97.25}  \\
\hline

\multirow{2}{*}{40} & MTL & 97.25 $\uparrow$ & 95.87 $\uparrow$ & 96.82 $\uparrow$ & 98.46 $\uparrow$ \\
  & \textcolor{gray}{STL} & \textcolor{gray}{96.74} & \textcolor{gray}{95.46} & \textcolor{gray}{96.23} & \textcolor{gray}{97.76}  \\
\hline

\multirow{2}{*}{80} & MTL & 97.29 $\uparrow$ & 96.30 $\uparrow$ & 97.18 $\uparrow$ & 98.60 $\uparrow$ \\
  & \textcolor{gray}{STL} & \textcolor{gray}{96.72} & \textcolor{gray}{95.84} & \textcolor{gray}{96.35} & \textcolor{gray}{97.71}  \\
\hline

\multirow{2}{*}{120} & MTL & 97.12 $\uparrow$ & 96.34 $\uparrow$ & 96.88 $\uparrow$ & 98.14 $\uparrow$ \\
  & \textcolor{gray}{STL} & \textcolor{gray}{97.01} & \textcolor{gray}{95.94} & \textcolor{gray}{96.60} & \textcolor{gray}{97.95}  \\
\Xhline{1.2pt}

\end{tabular}
}
\end{center}
\label{tab_eval_mtl}
\end{table}









\subsection{Experiment on resource consumption}

\subsubsection{\textbf{Compared with EKLAVYA}}

We compare \sys with EKLAVYA as shown in Table \ref{tabl_consumption_ek}.
About embedding, EKLAVYA uses 500 instructions as input compared with our 40 instructions (160 instruction words), and the vector is 256-dimensional compared with our 128-dimensional. Training represents the average time in seconds that the model takes to train one epoch. GT and CT represents the testing time on GPU and CPU for one function in milliseconds, respectively. 
Theoretically, the input matrix size of EKLAVYA is about \textbf{6}$\times$ than \sys. With MTL structure, \sys further improves efficiency. As a result, EKLAVYA is about \textbf{9.92}$\times$ longer than \sys in training time and \textbf{8}$\times$ longer both on GPU and CPU in testing time. 

We change embedding and size to further explore their effect on time consumption. Even with the same input as \sys, EKLAVYA still takes \textbf{1.56}$\times$ in training and \textbf{2.96}$\times$ in testing on GPU. With the same size, our embedding method leads to more time-consuming due to the use of instruction words, but our model also performs better when using instruction words. 


\begin{table}[!h]
\setlength\tabcolsep{2.5pt}
\begin{center}
\caption{Time consuming comparison. \textit{GT/CT} denotes testing time on GPU/CPU.}
\resizebox{0.49\textwidth}{!}{
\label{tabl_consumption_ek}
\begin{tabular}{c|c|c|c|c|c}
\Xhline{1.2pt}
\rowcolor{gray}
\textcolor{white}{\textbf{Structure}} & 
\textcolor{white}{\textbf{Embedding}} & 
\textcolor{white}{\textbf{Size}} & 
\textcolor{white}{\textbf{Training (s)}} & \textcolor{white}{\textbf{GT (ms)}} & \textcolor{white}{\textbf{CT (ms)}} \\ \Xhline{1.2pt}
STL & EKLAVYA & 500        & 942.01                & 17.88                      & 70.40
\\ \hline
STL & EKLAVYA & 40       & 132.72               & 6.56                      & 10.26
\\ \hline
STL & \sys & 40       & 216.05               & 7.87                      & 24.52
\\ \hline
\textbf{MTL} & \textbf{\sys}  & \textbf{40}        & \textbf{94.84 (9.9$\times$)}               & \textbf{2.20 (8.1$\times$)}                      & \textbf{8.39 (8.4$\times$)}                      \\ \Xhline{1.2pt}
\end{tabular}
}
\end{center}
\end{table}

The following experiments indicate that MTL saves resource consumption in all aspects.


\subsubsection{\textbf{Time consumption \& Efficiency}}
We record the time consumption of the previous experiments and calculate their efficiency, shown in Table \ref{tab_eval_eff}.
\textit{Time} represents the average time in seconds that the model takes to train one epoch. \textit{GPU} represents the average GPU usage percent during training.

\begin{table}[!h]
\caption{Model efficiency comparison.}
\begin{center}
\resizebox{0.49\textwidth}{!}{
\begin{tabular}{c|c|c|c|c}
\Xhline{1.2pt}
\rowcolor{gray}
\textcolor{white}{\textbf{Size}} & \textcolor{white}{\textbf{Structure}} & \textcolor{white}{\textbf{Time (s)}} & \textcolor{white}{\textbf{GPU (\%)}} & \textcolor{white}{\textbf{Efficiency (\%)}} \\

\Xhline{1.2pt}
\multirow{2}{*}{5} & MTL & 37.09 & 40.52 & 17.28 (2.09$\times$) \\
& \textcolor{gray}{STL} & \textcolor{gray}{122.27} & \textcolor{gray}{25.75} & \textcolor{gray}{8.27} \\
\hline

\multirow{2}{*}{10} & MTL & 40.19 & 58.73 & 14.91 (1.96$\times$) \\
& \textcolor{gray}{STL} & \textcolor{gray}{132.82} & \textcolor{gray}{35.00} & \textcolor{gray}{7.62} \\
\hline

\multirow{2}{*}{20} & MTL & 47.81 & 79.37 & 10.08 (2.31$\times$) \\
& \textcolor{gray}{STL} & \textcolor{gray}{139.09} & \textcolor{gray}{62.90} & \textcolor{gray}{4.37} \\
\hline

\multirow{2}{*}{40} & MTL & 94.84 & 78.58 & 5.21 (2.08$\times$) \\
& \textcolor{gray}{STL} & \textcolor{gray}{216.05} & \textcolor{gray}{71.25} & \textcolor{gray}{2.51} \\
\hline

\multirow{2}{*}{80} & MTL & 168.43 & 82.73 & 2.79 (2.10$\times$) \\
& \textcolor{gray}{STL} & \textcolor{gray}{380.44} & \textcolor{gray}{76.63} & \textcolor{gray}{1.33} \\
\Xhline{1.2pt}

\end{tabular}
}
\end{center}
\label{tab_eval_eff}
\end{table}

Obviously, the smaller size makes the training time shorter because it reduces the computational effort from the source. 
In addition, the MTL model takes less training time compared to \textcolor{gray}{STL} models of the same size. 
Furthermore, the time consumption of MTL increases less compared with STL as the size increases. For example, when size increases from 20 to 40, STL models need extra 76.96 minutes to train one epoch, but the MTL model only needs extra 47.03 minutes.
The above time-saving effect is attributed to the fact that MTL merges the task-specific layers of multiple STL models into the shared layer, thus avoiding duplicate computations. 
We find a similar phenomenon in efficiency, indicating that MTL also makes higher utilization of computational resources.

\begin{table}[!h]
\setlength\tabcolsep{14pt}
\caption{Model test time on both GPU and CPU.}

\begin{center}
\resizebox{0.49\textwidth}{!}{
\begin{tabular}{c|c|c|c}
\Xhline{1.2pt}
\rowcolor{gray}
\textcolor{white}{\textbf{Size}} & \textcolor{white}{\textbf{Structure}} & \textcolor{white}{\textbf{GPU (ms)}} & \textcolor{white}{\textbf{CPU (ms)}} \\
\Xhline{1.2pt}
\multirow{2}{*}{5} & MTL  & 1.47 & 2.18 \\
& \textcolor{gray}{STL} & \textcolor{gray}{5.64} & \textcolor{gray}{7.56} \\
\hline

\multirow{2}{*}{10} & MTL & 1.56 & 3.02 \\
& \textcolor{gray}{STL} & \textcolor{gray}{5.89} & \textcolor{gray}{10.12} \\
\hline

\multirow{2}{*}{20} & MTL & 1.78 & 4.84 \\
& \textcolor{gray}{STL} & \textcolor{gray}{6.58} & \textcolor{gray}{15.08} \\
\hline

\multirow{2}{*}{40} & MTL & 2.20 & 8.39 \\
& \textcolor{gray}{STL} & \textcolor{gray}{7.87}  & \textcolor{gray}{24.52} \\
\hline

\multirow{2}{*}{80} & MTL & 2.93 & 15.23 \\
& \textcolor{gray}{STL} & \textcolor{gray}{9.67} & \textcolor{gray}{42.64} \\
\Xhline{1.2pt}

\end{tabular}
}
\end{center}
\label{tab_eval_test_time}
\end{table}

The trained model needs to be used on different machines, which most likely do not have GPUs, so we use both GPU and CPU to invoke the model to make predictions on the test set and record the run time shown in Table \ref{tab_eval_test_time}. Here, GPU and CPU represent the average time spent predicting one function signature in milliseconds. 

As we can see, similar to the training, it takes less time testing with smaller input, and the time-saving effect of MTL also applies in testing both on CPU and GPU, so MTL is a CPU-friendly structure compared with STL. 
One important finding is that as the size increases (take MTL as an instance, from size 5 to 80), the time consumption of testing on the CPU ($6.98\times$) grows faster than that of the GPU ($1.99\times$), which further motivates us to use a shorter input to help CPU-only analysts improve efficiency.







\subsubsection{\textbf{Storage consumption}}
MTL reduces the number of models required to accomplish signature recovery, thus reducing the physical size that models require. In our experiment, MTL model only requires \textbf{32}MB while STL models requires \textbf{60}MB in total.
\section{discussion} \label{discussion}

\subsection{Representative work}
In the above, we compare the most relevant work EKLAVYA to our work. ReSIL is a function signature recovery system optimized for EKLAVYA, we will discuss the differences between ReSIL and \sys in the following paper.

\begin{figure}[!h]
\centerline{\includegraphics[width=0.4\textwidth]{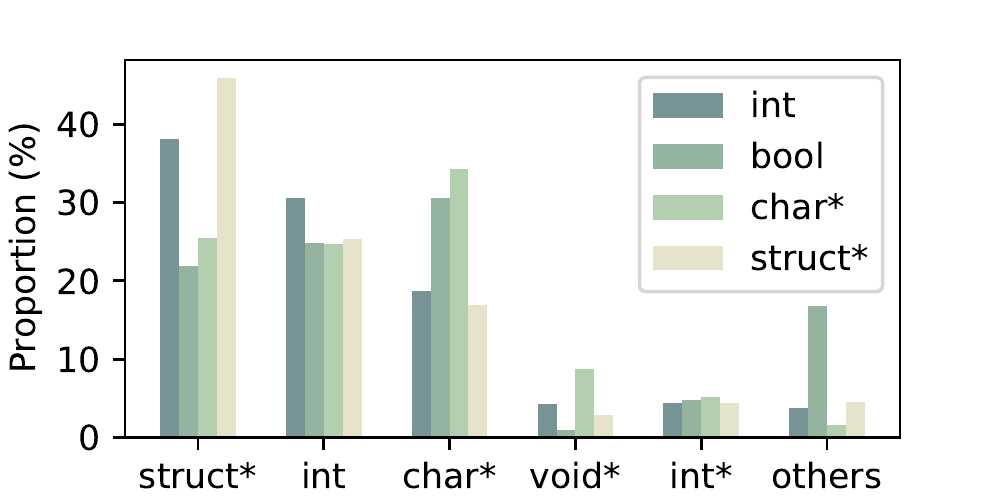}}
\caption{Distribution of parameter types under different return types.}
\label{discuss_rt_pt}
\end{figure}

EKLAVYA’s accuracy decreases when the inputs are optimized functions. ReSIL systematically discusses the reasons for the performance degradation and optimizes it. ReSIL improves the accuracy in inferring function signatures, for example, ReSIL improved the accuracy of recovery $PC$ from 84.8\% to 92.67\% at optimization level O1. 

ReSIL improves the accuracy by essentially using domain-specific knowledge. Inserting additional instructions to the input actually provides extra information from human knowledge, and removing unrelated instructions helps the model to focus more on useful features.
\begin{table}[!h]
\setlength\tabcolsep{15pt}
\caption{The effect of different location and input length of the instructions on the model accuracy.}
\begin{center}
\resizebox{0.3\textwidth}{!}{
\begin{tabular}{c|c|c}
\Xhline{1.2pt}
\rowcolor{gray}
\textcolor{white}{\textbf{Size}} & \textcolor{white}{\textbf{Location}} & \textcolor{white}{\textbf{RT (\%)}} \\
\Xhline{1.2pt}
\multirow{2}{*}{10} & Head & 67  \\
  & Tail & 72 \\
\hline
\multirow{2}{*}{20} & Head & 73 \\
  & Tail & 75 \\
\hline
\multirow{2}{*}{40} & Head & 74 \\
  & Tail & 76 \\
\Xhline{1.2pt}

\end{tabular}
}
\end{center}
\label{eval_bias}
\end{table}

Different from ReSIL, our work improves not only performance but also efficiency. We prefer to make the model more usable without introducing domain-specific knowledge. At the same time, ReSIL and our improvement approach are compatible, which means that both works can be applied simultaneously.

\subsection{Return type}
The return type, as one of the characteristics of the function, can also provide part of the pre-information for function signature recovery.
During exploratory data analysis, we find that the information of function return type also has spatial locality, clustered at the function tail. To verify the intuition, we do similar ablation experiments on return type and the results are shown in Table \ref{eval_bias}.

The result indicates that the instructions from the bottom of a function can provide more related information about the function return type.


In addition, we also find differences in the distribution of parameter types under different return types, as shown in Figure~\ref{discuss_rt_pt}. 
Obviously, functions with return type of \textit{struct*} employs more than 40\% of the parameters with type \textit{struct*}. The aforementioned information bias and distribution preference of return type may help improve function signature recovery or other security tasks in the future.

\section{related works}\label{related_works}

\subsection{Information recovery from stripped binaries}

Parameter type recovery in function signature recovery is essentially the process of recovering high-level semantic information (variable types) from stripped binaries, which can be seen as a special kind of task of type inference. Type inference can be basically divided into two categories, rule-based and machine learning-based. 

Lin et al. \cite{lin2010automatic} formulate rules based on expert knowledge for type inference.  \cite{caballero2009binary}, DIVINE \cite{balakrishnan2007divine}, TIE \cite{lee2011tie}, SecondWrite \cite{elwazeer2013scalable} also use analysis algorithms such as live variable analysis and manually formulate rules to infer variable tpyes. TypeArmor \cite{van2016tough} and $\tau$CFI \cite{muntean2018tau} use live variable analysis and heuristic methods to recover function signatures. Lin and Gao \cite{lin2021function} investigates the effect of optimization level on function signature recovery.

Some works introduce machine learning to the task and get some good results. BITY \cite{xu2017learning} uses the support vector machine (SVM) to classify type inference. TYPEMINER \cite{maier2019typeminer} uses both the Random Forest classifier and linear SVM to recover the type in multiple steps. CATI \cite{chen2020cati} uses CNNs, combined with assembly context as assistance to locate and infer variable types.

\subsection{Input for binary analysis}

Choosing an appropriate input is an important step for the success of binary analysis. However, due to the complexity and variety of binary analysis tasks, different inputs are used for different models and tasks to achieve the best results.

MECS \cite{10.1145/1014052.1014105} directly detect malicious executables with byte sequences. Rosenblum et al. \cite{10.5555/1620163.1620196} incorporates both idiom features and control flow structure features to identify function entry points. SMIT \cite{10.1145/1653662.1653736} extracts the function-call graph from a binary program and uses the graph matching algorithm to determine program similarity. ORIGIN \cite{10.1145/2001420.2001433} extracts significant features from binary programs and recovers provenance with the conditional random field. MutantX-S \cite{178990} extracts representative features from malware samples to cluster malware into families.TEDEM \cite{10.1145/2664243.2664269} automatically finds bugs with bug signatures. Pewny et al. \cite{7163056} lifts binary codes to the intermediate representation (IR) to obtain the semantics at a basic block level. Genius \cite{10.1145/2976749.2978370} converts the control-flow graphs into vectors and achieves realtime bug search. RENN \cite{8952186} learns memory alias dependencies with the binary encoding of instructions and memory region information. XDA \cite{pei2021xda} learns different contextual dependencies and disassembles with raw machine code. ASTERIA \cite{9505086} uses the abstract syntax tree (AST) to detect similarity with Tree-LSTM. jTrans \cite{10.1145/3533767.3534367} tokenizes raw assembly codes and embeds control flow information to detect binary code similarity.

\subsection{Multi-task learning}

Multi-task learning is a common effective machine learning architecture where multiple tasks are solved simultaneously. Due to domain information sharing between different tasks, the models are more generalized and robust compared with single-task learning \cite{10.1007/978-3-030-58536-5_10}. 

Many researchers try to improve the MTL performance by modifying the architecture. Misra et al. \cite{Misra_2016_CVPR} proposes ``cross-stitch'', a new sharing unit, to learn a combination of shared and task-specific representations. MRN \cite{NIPS2017_03e0704b} alleviates negative-transfer and under-transfer by jointly learning features and task relationships. Deep-AMTFL \cite{pmlr-v80-lee18d} prevents negative transfer by introducing an asymmetric autoencoder term. SNR \cite{Ma_Zhao_Chen_Li_Hong_Chi_2019} modularizes the shared low-level hidden layers into multiple layers and controls the connection of sub-networks to improve accuracy and maintain efficiency. 

MTL shows excellent results in many applications. Giri et al. \cite{7178925} uses MTL for speech recognition in reverberant environments. Isonuma et al. \cite{isonuma-etal-2017-extractive} addresses document summarization in the framework of MTL. Zou et al. \cite{10.1145/3178876.3186050} utilize MTL for web searching. Zhou et al. \cite{zhou-etal-2019-improving} improves the robustness of machine translation with MTL. MKM-SR  \cite{10.1145/3397271.3401098} use MTL for the session-based recommendation. Wang et al. \cite{9426045} proposed an MTL approach for code understanding. Xie et al. \cite{9462964} design an MTL method for code summarization. MTLFace \cite{Huang_2021_CVPR} alleviates the effect of age variation in face recognition with MTL. DeepCVA \cite{9678622} uses the MTL model to assess software vulnerabilities with better performance and less time.
\section{conclusion}\label{conclusion}

In this paper, we present an MTL-GRU model with selective inputs to accomplish function signature recovery and reduce resource consumption.
Based on the intuition of information bias, we used two selection strategies that are input length selection and input position selection. Experimental results verify our intuition that most information about function signature is gathered at the head of the function. In addition, motivated by the relationship between the function signature recovery tasks, we make the best use of the correlated information with multi-task learning.
As a result of the selective inputs and multi-task learning, our model improves recovery accuracy and greatly reduces resource consumption both in time and storage size.

\section*{Acknowledgment}

We sincerely thank the anonymous reviewers for their valuable comments helping us to improve this work. We are grateful to Zhongling He for his contributions and suggestions to the system construction and algorithm design in the early stage of the project. This work was supported in part by grants from the Chinese National Natural Science Foundation (61272078, 62032010, 62172201), the program B for Outstanding Ph.D. candidate of Nanjing University.

\bibliographystyle{ieeetr}
\bibliography{ref}

\end{document}